\documentclass[preprint,5p,sort&compress,times,twocolumn]{elsarticle}

\usepackage{amsthm, amsfonts, amssymb}	%%
\usepackage{epstopdf}        			%%
\usepackage[fleqn]{amsmath}				%%
\usepackage{graphicx}					%%
\usepackage{txfonts}					%%
\usepackage{hyperref}					%%
\usepackage{lineno,color,comment}

\modulolinenumbers[5]					%%
\begin{document}

\begin{frontmatter}
\title{Algebraic structures and deformed Schr\"{o}dinger equations from groups entropies}
%\title{Deformed algebra and calculus from formal Lazard group law}

\author[ufba,cnpq]{Ignacio S. Gomez\corref{cor1}}
\ead{nachosky@fisica.unlp.edu.ar}
%\author[if-sertao]{Bruno G. da Costa}
%\ead{bruno.costa@ifsertao-pe.edu.br}
\author[ufba]{Ernesto P. Borges}
\ead{ernesto@ufba.br}
%\author[iflp,unlp]{M. Portesi}
%\ead{portesi@fisica.unlp.edu.ar}

\cortext[cor1]{Corresponding author}
\address[ufba]{Instituto de F\'{i}sica, Universidade Federal da Bahia,
         Rua Barao de Jeremoabo, 40170-115 Salvador--BA, Brasil}
\address[cnpq]{Conselho Nacional de
Desenvolvimento Cient\'ifico e Tecnol\'ogico (CNPq), Edif\'icio Santos Dumont,
71.605-001, Lago Sul, Brasília - DF, Brasil}

 %            Sert\~ao Pernambucano,
  %           BR 407, km 08, 56314-520 Petrolina, Pernambuco, Brasil}
%\address[iflp]{IFLP, CONICET, UNLP, Boulevard 113 e/63 y 64, 1900 La Plata, Argentina}
%\address[unlp]{Facultad de Ciencias Exactas, Universidad
%	Nacional de La Plata, C.C. 67, 1900 La Plata, Argentina}

\begin{abstract}
Motivated by the group entropy theory,
in this work we generalize the algebra of real numbers (that we called
$G$-algebra),
%based on the Lazard formal group law,
%which is related to the nonextensive statistics inspired
%by the group entropy theory.
%By means of the $G$-algebra
from which we develop an associated $G$-differential calculus.
%, is also presented.
% from which
%their generalized derivative and integral, are presented.
Thus, the algebraic structures corresponding to
the Tsallis and Kappa statistics
% $q$-algebra and the Kaniadakis ones,
%, along with their associated calculus,
are obtained
as special cases when the Tsallis and Kappa
group classes are chosen. We employ the $G$-algebra to formulate a
generalized $G$-deformed Schr\"{o}dinger equation and we
illustrate it with the infinite potential well, where the
effective mass is related with the $G$-algebra structure
and the $q$-deformed (standard) Schr\"{o}dinger equation results an special case
for the Tsallis (Boltzmann-Gibbs) group class.
The
non-uniform zeros spacing of the $G$-deformed eigenfunctions
is expressed in terms of the generalized sum of the $G$-algebra.
 %We illustrate the formalism with $G-$deformed
%versions of the (non-relativistic) Schr\"{o}dinger equation corresponding
%to the Tsallis and Kaniadakis group classes and we find the autofunctions for
%case of the unidimensional infinite potential well.
\end{abstract}


\end{frontmatter}

\nolinenumbers

\section{Introduction}
\label{intro}

The achievement of nonextensive statistics in describing multiple phenomena (where the
application of the traditional
Boltzmann-Gibbs (BG) statistical mechanics have shown some problems)
as
anomalous diffusion \cite{Tsa96},
long-range interactions \cite{Kal14},
plasmas \cite{Guo12}, quantum tunneling \cite{Aqu17},
cold atoms \cite{Dou06}, onset of chaos \cite{Tir16},
quantum chaos scales \cite{Gom-Bor18}, and measure theory \cite{Gomez-RPM-2019}
among others, motivated the
construction of several approaches of generalized statistical mechanics from which the
foundations of the theory have been extended \cite{Tsa88,tsallis-book}.
In these advances, many mathematical frameworks associated with generalized statistics
have been proposed \cite{Abe98,Kan01,Kan02,Niv03,Bor04}.
Concerning with the nonextensive Tsallis and
Kaniadakis entropies, their associated
algebraic structures, called the $q$-algebra \cite{Niv03,Bor04}
and the $\kappa$-algebra \cite{Kan01,Kan02},
were formulated. In particular, the $q$-algebra
has been employed to characterize position-dependent
mass systems \cite{Richstone_1982,vonroos_1983,Alhaidari-2004,Ranada-2016}
from a
framework doted with a deformed space
and a deformed
derivative, where the functional form of the
effective mass is related with the deformation inherited
from the algebraic structure
\cite{CostaFilho-Almeida-Farias-AndradeJr-2011,
CostaFilho-Alencar-Skagerstam-AndradeJr-2013,
Costa-Borges-2014,Costa-Borges-2018,Costa-Gomez-2018}.

To provide a unified scenario for generalized entropies in nonextensive statistics
from a group theory viewpoint,
%an interesting mathematical framework
%has been recently developed, the so-called
the group entropy approach has been developed
\cite{Tem11,Tem15,Tem15bis,Tem16,Tem17,Tem19}.
In this framework, the entropy functional is proposed in function of the
class of interactions that are considered and it defines a group entropy
that satisfies the three
Shannon-Kinchin (SK) axioms and the composability action.
Each group entropy is expressed in terms
of a formal group element, being the formal group law a generalized additivity.
The group entropies have been recently
studied as generalized information measures in information geometry
%for characterizing correlations in curved statistical manifolds
% are information measures since their associated
%group Kullback-Leibler divergence and group Fisher metric allow to
%characterize states and curved statistical manifolds
\cite{Tem19}.   %
From group entropy approach the generalized logarithm and its inverse
are derived: the group logarithm and the group exponential. When the group entropy
corresponds to the BG class, all the standard definitions
are recovered.

Inspired by the $q$-algebra
and the $\kappa$-algebra, the goal of this letter is to propose a
generalized algebraic structure, that we called $G$-algebra,
based on the formal group law, the group
logarithm ($G$-logarithm) and the group exponential ($G$-exponential).
Thus, the present proposal allows to obtain several deformed algebraic structures
in a unified way,
in terms of the group entropy considered.
The $q$-algebra and the
$\kappa$-algebra are obtained as special cases for the Tsallis and Kaniadakis classes.
From the $G$-algebra we construct its associated differential
calculus, that we called \emph{$G$-calculus},
provided with the $G$-differential, the $G$-derivative, and the $G$-integral.

The letter is structured as follows.
In Section~2 we review the concept of
group entropy, along
with the Lazard formal group and
some properties.
With the help of the formal group law, the
$G$-logarithm and the $G$-exponential in Section~3 we
present the $G$-algebra; where the operations of the real numbers
(sum, substraction, product and division)
are generalized.
%in terms of the group entropy chosen.
Next, we present the $G$-differential calculus that can be derived from the $G$-algebra,
where the $G$-differential, the $G$-derivative and the $G$-integral are developed.
We show that the usual algebra, the $q$-algebra and the $\kappa$-algebra
are obtained when the BG, the Tsallis and the Kaniadakis classes are
considered, respectively.
Then, we present a $G$-deformed Schr\"{o}dinger equation
and we study the corresponding versions to the Tsallis
and the Kaniadakis classes.
We solve
them
%explicitly
for
%the case of
the unidimensional
infinite potential well and we obtain the
%The
corresponding effective masses, that
are related with their
%deformed
algebraic structures according to
\cite{Costa-Borges-2014,Costa-Borges-2018}.
Finally, in Section~4 some conclusions and perspectives are outlined.

\section{Review of group entropy theory}\label{sec prelim}

In this section we give the notions and concepts necessary for the development
of the paper. We follow the exposition given in the
Refs. \cite{Tem11,Tem15,Tem15bis,Tem16,Tem17,Tem19}.

The group entropy approach is a construction inspired
from generalizations of the notion of entropy
that are compatible with the SK axioms, by making use of the
formal group theory, thus giving a connection between
entropic measures and formal group theory.
Given a entropy $S_n$ ($n\in\mathbb{N}$)
defined over the set of probability
distributions $\{p=(p_1,\ldots,p_n): \sum_{i=1}^n p_i=1 \ , \ p_j\geq 0\}$ the SK axioms establishes:
\begin{itemize}
  \item[(I)] the continuity of the functional $S_n[p]$,

  \item[(II)] the maximum of $S_n$ is attained by the uniform distribution,

  \item[(III)] $S_n$ is expandable to $S_{n+1}$ by adding events of zero probability,

  \item[(IV)] \textbf{Composability Axiom}: $S_{nm}[pq]=S_{n}[p]+S_m[q|p]$ where
$S_m[q|p]=\sum_{k=1}^n p_k S_m[q]$, $(pq)_{ik}=p_k q_{i|k}$ for all $i=1,\ldots,m$ and
$k=1,\ldots,n$.
\end{itemize}
It can be shown that the Boltzmann-Gibbs entropy
$S[p]=\sum_{i=1}^n p_i \log(p_i^{-1})$ is the only one that satisfies the SK axioms.
In this axiomatical characterization of the Boltzmann-Gibbs entropy by means of first principles
the composability axiom turns to be essential.

With the aim to generalize the composability axiom for deriving more entropic forms,
%in the works []
the notion of group entropy was introduced. A \emph{group entropy} is
a function satisfying the (I)--(III) SK axioms along with following generalized
composable notion.
If there exist an entropy $S$ and a continuous real function $\Phi(x,y)$ such that
\begin{itemize}
  \item[(C1)] \emph{Composability:} If $A$ and $B$ are two statistically independent
  systems provided with probability distribution $\{p_i\}_{i=1}^n$ and $\{q_j\}_{j=1}^m$
  then $S(A \cup B)=\Phi(S(A),S(B))$.

  \item[(C2)] \emph{Symmetry:} $\Phi(x,y)=\Phi(y,x)$.

  \item[(C3)] \emph{Associativity:} $\Phi(x,\Phi(y,z))=\Phi(\Phi(x,y),z)$.

  \item[(C4)] \emph{Null-composability:} $\Phi(x,0)=x$.
\end{itemize}
then we say that $S$ is \emph{strongly composable}.
The importance of this definition is that C1--C4 generalize the
(IV) SK axiom in such a way that a uniqueness theorem for
an entropic form (which depends on $\Phi$)
satisfying the (I)--(III) SK axioms and the C1--C4 conditions,
can be guaranteed \cite{Tem15bis,Tem17}.

A \emph{formal group law} defined over
the real numbers $\mathbb{R}$ is any power series $\Phi(x,y)$
with real coefficients satisfying C2--C4. The polynomial
$\Phi(x,y)$ can be expressed by the \emph{Lazard formal group law}
\begin{eqnarray}\label{formal-group-law}
\Phi(x,y)=G(G^{-1}(x)+G^{-1}(y))=x \oplus_G y
\end{eqnarray}
where: $\oplus_G$ denotes the generalized sum associated
with the formal group structure,
\begin{eqnarray}\label{group-G}
G(t)=t+\sum_{k=1}^{\infty} a_k \frac{t^{k+1}}{k+1}
\end{eqnarray}
is the formal group exponential and
\begin{eqnarray}\label{group-G}
F(s)=G^{-1}(s)=t+\sum_{i=1}^{\infty} b_i \frac{s^{i+1}}{i+1}
\end{eqnarray}
is the formal group logarithm, which
satisfy $F(G(t))=t$, $G(F(s))=s$, $b_i\in \mathbb{Z}$, and
$a_1=-b_1, a_2=\frac{3}{2}b_1^2-b_2,\ldots$
Using these expressions, the group logarithm $\log_G(x)$ and the
group exponential $\exp_G(x)$ are defined as
\begin{eqnarray}\label{G-logarithm}
\log_G(x)=G(\log x) \quad \quad \quad \textrm{(\emph{group logarithm})}
\end{eqnarray}
and
\begin{eqnarray}\label{G-exponential}
\exp_G(x)=\exp(G^{-1}(x)) \quad \quad \quad \textrm{(\emph{group exponential})}.
\end{eqnarray}
Postulating the reasonable requirement for the group logarithm
\begin{eqnarray}\label{G-logarithm-product}
\log_G(x \otimes_G y)= \log_G (x) + \log_G (y)
\end{eqnarray}
that constitutes the definition of the product $\otimes_G$ derived
from the formal group structure,
and using that $\log_G (x)$ and $\exp_G (x)$ are one the inverse
of the other
we can obtain
\begin{eqnarray}\label{G-product}
x \otimes_G y= \exp_G(\log_G (x) + \log_G (y)).
\end{eqnarray}
From this equation and the definitions of the group logarithm and
the group exponential
it follows that
\begin{eqnarray}\label{G-property-exponential-1}
\exp_G(x+y)=\exp_G (x)\otimes_G\exp_G (y),
\end{eqnarray}
\begin{eqnarray}\label{G-property-exponential-2}
\exp_G(x)\exp_G(y)=\exp_G (x \oplus_G y),
\end{eqnarray}
and
\begin{eqnarray}\label{G-property-logarithm}
\log_G(xy)=\log_G (x)\oplus_G \log_G(y).
\end{eqnarray}
One of the main advantages of the group entropy approach is that it allows to
express the additivity property in a unified way. For instance, if we choose
$\Phi_{\mathcal{BG}}(x,y)=x+y$ and
$\Phi_{q}(x,y)=x+y+(1-q)xy$ (with $q$ a real continuous parameter)
we recover the usual
additivity and the non-additivity of the $q$-Tsallis entropies,
which correspond to the group entropies
\begin{eqnarray}\label{BG-class}
G_{\mathcal{BG}}(t)=t \quad \quad \quad \quad \quad \textrm{(\emph{BG-class})}
\end{eqnarray}
and
\begin{eqnarray}\label{Tsallis-class}
G_{q}(t)=\frac{\exp((1-q)t)-1}{1-q} \quad \quad \quad \textrm{(\emph{Tsallis class})}
\end{eqnarray}
In such cases the group logarithm and the group exponential
become into the standard ones for the Boltzmann-Gibbs
class, and into the $q$-logarithm
\begin{eqnarray}\label{q-logarithm}
\log_q x =\frac{x^{1-q}-1}{1-q}
\end{eqnarray}
along with the $q$-exponential
\begin{eqnarray}\label{q-exponential}
\exp_q^x=[1+(1-q)x]_{+}^{\frac{1}{1-q}} \quad , \quad ( \ x\in [-1/(1-q),\infty) \ )
\end{eqnarray}
for the Tsallis class.
Analogously, the deformed logarithm and exponential functions associated to the
Kaniadakis class and the Abe class can be obtained from their group entropies
\begin{eqnarray}\label{Kaniadakis-class}
G_{\mathcal{K}}(t)=\frac{\exp(\sigma t)-\exp(-\sigma t)}{2\sigma}
\quad \quad \quad \textrm{(\emph{Kaniadakis class})}
\end{eqnarray}
and
\begin{eqnarray}\label{Abe-class}
G_{a,b}(t)=\frac{\exp(at)-\exp(bt)}{a-b}
\quad \quad \quad \textrm{(\emph{Abe class})}
\end{eqnarray}
with $\sigma,a,b\in \mathbb{R}$.

\section{Algebraic structure and calculus associated to
the groups entropies}

In order to complete
the algebraic structure corresponding to the
Lazard formal group law, from Eqns. \eqref{formal-group-law} and
\eqref{G-product} (that we call from now on $G$-sum and $G$-product)
we need to define their neutral elements,
along with the $G$-substraction, the $G$-division and their inverse elements.
Given $x,y \in \mathbb{R}$ it is not difficult to see that the ``$G$-algebra" desired
is given by the
following generalized operations in terms of the group entropy $G(t)$:
\begin{itemize}
  \item [(a)] \emph{G-sum:} $x\oplus_G y=G(G^{-1}(x)+G^{-1}(y))$
  \item [(b)] \emph{G-substraction:} $x\ominus_Gy=G(G^{-1}(x)-G^{-1}(y))$
  \item [(c)] \emph{G-product:} $x \otimes_G y=\exp_G(\log_G(x)+\log_G(y))$
  \item [(d)] \emph{G-division:} $x \oslash_G y=\exp_G(\log_G(x)-\log_G(y))$
\end{itemize}
together with the neutral and inverse elements:
\begin{itemize}
  \item [(e)] \emph{neutral element of G-sum:} $x\oplus_G 0=0\oplus_G x=x$
  \item [(f)] \emph{inverse of G-sum:} $\ominus_G x=G(-G^{-1}(x))$
  \item [(g)] \emph{neutral element of G-product:}
  $x \otimes_G 1=1 \otimes_G x=x$
  \item [(h)] \emph{inverse of G-product:} $\oslash_G x=\exp_G(-\log_G(y))$
\end{itemize}
which naturally become into the standard operations for the
BG-class $G(t)=t$.
From the list (a)-(h) we have that the $G$-sum and the $G$-product
are associative, i.e. $x\oplus_G (y \oplus_G z)=(x\oplus_G y) \oplus_G z$ and
$x\otimes_G (y \otimes_G z)=(x\otimes_G y) \otimes_G z$
but the $G$-sum is not distributive
neither in relation to the usual product
($a(x\oplus_G y)\neq ax \oplus_G ay$) nor to the $G$-product
($a\otimes_G (x\oplus_G y)\neq (a \otimes_G x) \oplus_G (a \otimes_G y)$)
(see for instance \cite{Bor04}).
The $G$-sum and the $G$-product allow to generalize the notion of integer
and power of a number in the following way:
\begin{itemize}
  \item [(i)] \emph{$G$-integers:} $n_G=G(nG^{-1}(1))$ for all $n\in \mathbb{Z}$.
  \item [(j)] \emph{$G$-powers of $x$:} $x^{{G}_n}=\exp_G(n\log_G (x))$
  for all $n\in \mathbb{Z}$.
\end{itemize}
The $G$-integers (that we denote by $\mathbb{Z}_G$)
satisfy
\begin{eqnarray}\label{q-number-integers}
(n+m)_G=n_G \oplus_G m_G
\end{eqnarray}
for all $n,m\in\mathbb{Z}$.
Then, the map $n\mapsto n_G$ is a group isomorphism between the
integers with the usual sum $(\mathbb{Z},+)$ and the
$G$-integers with the $G$-sum $(\mathbb{Z}_G,\oplus_G)$, i.e.
$(\mathbb{Z},+)\simeq (\mathbb{Z}_G,\oplus_G)$.

Considering the Tsallis class $G_q(t)$,
the $q$-logarithm
and the $q$-exponential have motivated
the $q$-deformation of a number $x$
\cite{Tsa88,Niv03,Bor04},
given by
\begin{eqnarray}\label{q-number}
x_q=\log (\exp_q(x))=G_q^{-1}(x)
\quad \quad (q-\textrm{\emph{deformation of}}x)
\end{eqnarray}
Following this recipe,
we define the $G$-deformation of x as
\begin{eqnarray}\label{G-number}
x_G=\log (\exp_G(x))=G^{-1}(x)
\quad \quad (G-\textrm{\emph{deformation of}}x)
\end{eqnarray}
From this definition we have the property:
\begin{eqnarray}\label{G-number-property}
%\begin{array}{ccc}
(x \oplus_G y)_G & = & x_G+y_G
%\nonumber\\
%x \otimes_G y & = & x_Gy_G
%\end{array}
\end{eqnarray}
for all $x,y\in \mathbb{R}$. The property
\eqref{G-number-property} defines an isomorphism between
the groups $(\mathbb{R},\oplus_G)$ and $(\mathbb{R}_G,+)$
where $\mathbb{R}_G$ stands for the $G$-real numbers, i.e.
$\mathbb{R}_G:=\{x_G=G^{-1}(x): x\in\mathbb{R}\}$.
Other deformation by means of the group entropy $G$ can be
proposed as follows, that we called $G$-dual deformation
\begin{eqnarray}\label{G-dual-number}
\widetilde{x_{G}}=\log_G(\exp(x))=G(x)
\quad (G-\textrm{\emph{dual deformation of}}x)
\end{eqnarray}
from which we have
\begin{eqnarray}\label{G-dual-number-property}
\widetilde{(x \oplus_G y)_{G}} & = & \widetilde{x_{G}}+\widetilde{y_{G}}
\end{eqnarray}
for all $x,y\in \mathbb{R}$.
This defines a group isomorphism between
$(\mathbb{R},\oplus_G)$ and $(\widetilde{\mathbb{R}_G},+)$
where $\widetilde{\mathbb{R}_G}:=
\{\widetilde{x_G}=G(x): x\in\mathbb{R}\}$
stands for the $G$-dual real numbers. Naturally, for all $x\in\mathbb{R}$
we also have
\begin{eqnarray}\label{G-G-dual-property}
(\widetilde{x_G})_G=\widetilde{(x_G)_G}=x.
\end{eqnarray}
Thus, the following isomorphisms result
\begin{eqnarray}\label{isomorphisms}
(\mathbb{R}_G,+)\simeq (\mathbb{R},\oplus_G)\simeq (\widetilde{\mathbb{R}_G},+)
\end{eqnarray}
which will play an important role for constructing a $G$-deformed calculus.

\subsection{$G$-calculus and its dual structure}

From the definitions of the $G$-deformation and its dual one we can derive
their corresponding differentials by means of the
identity $G(G^{-1}(x))=G^{-1}(G(x))=x$.
We obtain
\begin{eqnarray}\label{G-G-dual-differentials-1}
dx=G^{\prime}(G^{-1}(x))dx_G
\end{eqnarray}
and
\begin{eqnarray}\label{G-G-dual-differentials-2}
dx=\frac{1}{G^{\prime}(x)}d \widetilde{x_G}
\end{eqnarray}
from which follows
\begin{eqnarray}\label{G-G-dual-differentials-3}
\frac{d \widetilde{x_G}}{dx_G}=G^{\prime}(G^{-1}(x))G^{\prime}(x)
\end{eqnarray}
that establishes the relationship between the differential structures
of $(\mathbb{R}_G,+)$ and
$(\widetilde{\mathbb{R}_G},+)$.
Given a function $f:\mathbb{R}\rightarrow\mathbb{R}$,
from Eqns. \eqref{G-G-dual-differentials-1}--\eqref{G-G-dual-differentials-3}
we define the $G$-derivative
\begin{eqnarray}\label{G-derivative}
D_G f(x) = \frac{df}{dx_G}=G^{\prime}(G^{-1}(x))\frac{df}{dx}
\end{eqnarray}
and the corresponding one to the $G$-dual deformation
\begin{eqnarray}\label{G-derivative}
\widetilde{D_G} f(x) = \frac{df}{d \widetilde{x_G}}
=\frac{1}{G^{\prime}(x)}\frac{df}{dx}
\end{eqnarray}
which implies
\begin{eqnarray}\label{G-deformed-derivatives}
\frac{D_G f(x)}{\widetilde{D_G} f(x)}=G^{\prime}(G^{-1}(x))G^{\prime}(x).
\end{eqnarray}
Thus, the associated $G$-integral (noted by $I_G=\int_{(G)}dx_G$) is
\begin{eqnarray}\label{G-integral}
I_G f(x) = \int_{(G)}f(x)d x_G = \int f(x)\frac{dx}{G^{\prime}(G^{-1}(x))}
\end{eqnarray}
along with its dual version
\begin{eqnarray}\label{G-dual-integral}
\widetilde{I_G} f(x) = \int_{\widetilde{(G)}}
f(x)d \widetilde{x_G} =
\int f(x)G^{\prime}(x)dx.
\end{eqnarray}
Naturally, we have
\begin{eqnarray}\label{G-dual-integral}
\int_{(G)}D_G f(x)dx_G = \int_{\widetilde{(G)}}
\widetilde{D_G} f(x)d\widetilde{x_G}=f(x)+\textrm{constant}
\end{eqnarray}
which is an expression of the fundamental calculus theorem
within the $G$-deformed structure.

\subsection{Some algebraic structures as special cases}

We begin illustrating the relevance of the formalism presented
by studying the algebra and calculus of
two particular classes: the Tsallis and the Kaniadakis
group entropies (Eqns. \eqref{Tsallis-class} and \eqref{Kaniadakis-class}).
We omit the Abe class since the inverse
of the group entropy \eqref{Abe-class} has not an explicit formula.

\subsubsection{$q$-Algebra and calculus}

Considering the Tsallis class and replacing
the group entropy \eqref{Tsallis-class}, its inverse
together with the
$q$-exponential and the $q$-logarithm
in the list (a)-(h) we obtain the full structure
of the $q$-algebra \cite{Niv03,Bor04}:
\begin{itemize}
  \item[$(a_q)$] $q$-\emph{sum:} $x\oplus_q y=x+y+(1-q)xy$
  \item[$(b_q)$] $q$-\emph{substraction:} $x\ominus_q y=\frac{x-y}{1+(1-q)y}$ \quad $(y\neq \frac{-1}{1-q})$
  \item[$(c_q)$] $q$-\emph{product:} $x\otimes_q y=\left[x^{1-q}+y^{1-q}-1\right]_+^{\frac{1}{1-q}}$
  \item[$(d_q)$] $q$-\emph{division:} $x\oslash_q y=\left[x^{1-q}-y^{1-q}+1\right]_+^{\frac{1}{1-q}}$
  \item[$(e_q)$] \emph{neutral element of $q$-sum:} $x\oplus_q 0=0\oplus_q x=x$
  \item[$(f_q)$] \emph{inverse of $q$-sum:} $\ominus_q x=\frac{-x}{1+(1-q)x}$ \quad $(x\neq \frac{-1}{1-q})$
  \item[$(g_q)$] \emph{neutral element of $q$-product:} $x\otimes_q 1=1\otimes_q x=x$
  \item[$(h_q)$] \emph{inverse of $q$-product:} $1\oslash_q x=\left[2-x^{1-q}\right]_+^{\frac{1}{1-q}}$
\end{itemize}
along with the $q$-integers and the $q$-powers for all $n\in \mathbb{Z}$
\begin{itemize}
  \item[$(i_q)$] \emph{$q$-integers:} $n_q=\frac{1}{1-q}([2-q]^n-1)$
  \item[$(j_q)$] \emph{$q$-powers of $n$:} $x^{\otimes_{q}^n}=
  \left[nx^{1-q}-(n-1)\right]_+^{\frac{1}{1-q}}$.
\end{itemize}
Moreover, the $G$-derivative and the $G$-integral become
into the $q$-derivative
\begin{eqnarray}\label{q-derivative}
D_q f(x)=\frac{df}{dx_q}=(1+(1-q)x)\frac{df}{dx}
\end{eqnarray}
and the $q$-integral
\begin{eqnarray}\label{q-integral}
I_q f(x) = \int_{(q)}f(x)d x_q = \int f(x)\frac{dx}{(1+(1-q)x)}.
\end{eqnarray}
We also obtain the $q$-derivative and
the $q$-integral associated to the dual $q$-deformation:
\begin{eqnarray}\label{q-dual-derivative}
\widetilde{D_q} f(x)=\frac{df}{d\widetilde{x_q}}=\exp(-(1-q)x)\frac{df}{dx}
\end{eqnarray}
and
\begin{eqnarray}\label{q-dual-integral}
\widetilde{I_q} f(x) = \int_{\widetilde{(q)}}f(x)d \widetilde{x_q} =
\int f(x)\exp((1-q)x)dx.
\end{eqnarray}

\subsubsection{$\kappa$-algebra and calculus}

For the case of the Kaniadakis class (Eq. \eqref{Kaniadakis-class}),
by letting $\kappa=\sigma$,
the definitions (a)-(h) provide the full structure
of the $\kappa$-algebra \cite{Kan01,Kan02}:
\begin{itemize}
  \item[$(a_{\mathcal{K}})$] $\kappa$-\emph{sum:} $x\oplus_\kappa y=
  x\sqrt{1+\kappa^2y^2}+y\sqrt{1+\kappa^2x^2}$
  \item[$(b_{\mathcal{K}})$] $\kappa$-\emph{substraction:}
  $x\ominus_\kappa y=
  x\sqrt{1+\kappa^2y^2}-y\sqrt{1+\kappa^2x^2}$
  \item[$(c_{\mathcal{K}})$] $\kappa$-\emph{product:} \\
\noindent   $x\otimes_{\kappa}y=$ $\exp\left(\frac{1}{\kappa}
  (\textrm{arcsinh}((x^{\kappa}+y^{\kappa}-x^{-\kappa}-y^{-\kappa})/2)\right)$
  \item[$(d_{\mathcal{K}})$] $\kappa$-\emph{division:} \\
\noindent   $x\oslash_{\kappa}y=$ $\exp\left(\frac{1}{\kappa}
  (\textrm{arcsinh}((x^{\kappa}-y^{\kappa}-x^{-\kappa}+y^{-\kappa})/2)\right)$
    \item[$(e_{\mathcal{K}})$] \emph{neutral element of $\kappa$-sum:}
    $x\oplus_{\kappa} 0=0\oplus_{\kappa} x=x$
  \item[$(f_{\mathcal{K}})$] \emph{inverse of $\kappa$-sum:}
  $\ominus_{\kappa} x=\frac{-x}{1+(1-q)x}$ \quad $(x\neq \frac{-1}{1-q})$
  \item[$(g_{\mathcal{K}})$] \emph{neutral element of $\kappa$-product:} $x\otimes_\kappa 1=
  1\otimes_\kappa x=x$
  \item[$(h_{\mathcal{K}})$] \emph{inverse of $q$-product:}
  $1\oslash_\kappa x=x^{-1}$
\end{itemize}
where we have used that the $\kappa$-logarithm is
\begin{eqnarray}\label{kappa-logarithm}
\log_\kappa (x)=\frac{x^{\kappa}-x^{-\kappa}}{2\kappa}  \quad \quad (x\neq0)
\end{eqnarray}
and the $\kappa$-exponential is
\begin{eqnarray}\label{kappa-exponential}
\exp_\kappa (x)=\left[\kappa x+\sqrt{\kappa^2x^2+1} \right]_+^{\frac{1}{\kappa}}.
\end{eqnarray}
For all $n\in\mathbb{Z}$ the ${\mathcal{K}}$-integers and the
${\mathcal{K}}$-powers are given by
\begin{itemize}
  \item[$(i_{\mathcal{K}})$] \emph{$\kappa$-integers:}
  $n_\kappa=\frac{1}{\kappa}\sinh\left(n \textrm{arcsinh}(\kappa x)\right)$
  \item[$(j_{\mathcal{K}})$] \emph{$\kappa$-powers of $n$:} \\
\noindent  $x^{\otimes_{\kappa}^n}=\left[n\sinh(\kappa \log x)+
  \sqrt{n^2 \sinh^2(\kappa \log x)+1}\right]_+^{\frac{1}{\kappa}}$.
\end{itemize}
The ${\mathcal{K}}$-derivative and the ${\mathcal{K}}$-integral result
\begin{eqnarray}\label{kappa-derivative}
\mathcal{D}_\kappa f(x)=\frac{df}{dx_\kappa}=
\sqrt{(\kappa x)^2+1}\frac{df}{dx}
\end{eqnarray}
and the $\kappa$-integral
\begin{equation}\label{kappa-integral}
I_\kappa f(x) = \int_{(\kappa)}f(x)d x_\kappa =
\int f(x)\frac{1}{\sqrt{(\kappa x)^2+1}}dx
\end{equation}
We also have the ${\mathcal{K}}$-derivative and
the ${\mathcal{K}}$-integral associated to the dual
Kaniadakis deformation, expressed by
\begin{eqnarray}\label{kappa-dual-derivative}
\widetilde{D_\kappa} f(x)=\frac{df}{d\widetilde{x_\kappa}}=
\frac{1}{\cosh(\kappa x)}\frac{df}{dx}
\end{eqnarray}
and
\begin{eqnarray}\label{kappa-dual-integral}
\widetilde{I_\kappa} f(x) =
\int_{\widetilde{(\kappa)}}f(x)d \widetilde{x_\kappa} =
\int f(x)\cosh(\kappa x)dx.
\end{eqnarray}

%We illustrate the formalism presented with some examples.
%We begin by obtaining the $q$-algebra and the $\Kappa$-algebra
%as special cases when the group entropies correspond
%to the

\subsection{$G$-Deformed version of the
Schr\"{o}dinger equation}
With the aim to formulate a nonrelativistic Schr\"{o}dinger
in the context of group entropy theory, we propose a generalized
translation operator, given by
\begin{eqnarray}\label{G-translation-operator}
\hat{\mathcal{T}}_G(\varepsilon)|x\rangle=
|x \oplus_G \varepsilon\rangle=|G(G^{-1}(x)+G^{-1}(\varepsilon))\rangle
\end{eqnarray}
where the nonlinearity of the infinitesimal displacement $\varepsilon$ depends on the group entropy
$G$. Analogously to the standard case $G(t)=t$, it can be shown that the
operator $\hat{\mathcal{T}}_G(\varepsilon)$ has the following associated position-dependent
momentum
\begin{eqnarray}\label{G-momentum-operator}
\hat{p}_G=G^{\prime}(G^{-1}(\hat{x}))\hat{p}
\end{eqnarray}
being $\hat{p}=-i\hbar \partial/\partial x$ the standard momentum operator.
With the aim of guaranteeing
classical analogues, real eigenvalues and a orthonormal basis, we redefine $\hat{p}_G$ as
\begin{eqnarray}\label{G-momentum-operator-hermitic}
\hat{p}_G=\frac{1}{2}[G^{\prime}(G^{-1}(\hat{x}))\hat{p}+
\hat{p}(G^{\prime}(G^{-1}(\hat{x}))]
\end{eqnarray}
that is Hermitian. From the well-known commutator
$[\hat{x},\hat{p}]=
i\hbar\hat{1}$ (being $\hat{1}$ the identity operator) and using that
for all analytical
function $f(x)$ is
$[f(\hat{x}),\hat{p}]=
i\hbar f^{\prime}(\hat{x})$, we have that
\begin{eqnarray}
\begin{array}{ll}
\label{G-conjugated-operators}
& \hat{x}_G=G^{-1}(\hat{x}) \\
& \hat{p}_G=\frac{1}{2}[G^{\prime}(G^{-1}(\hat{x}))\hat{p}+
\hat{p}(G^{\prime}(G^{-1}(\hat{x}))]
\end{array}
\end{eqnarray}
satisfy
\begin{eqnarray}\label{G-commutator}
[\hat{x}_G,\hat{p}_G]=i\hbar\hat{1}.
\end{eqnarray}
The relationship \eqref{G-commutator}
shows that $\{\hat{x},\hat{p}\}\mapsto \{\hat{x}_G,\hat{p}_G\}$
is a canonical transformation and constitutes a ``natural" generalization
of the conjugated operators $\hat{x},\hat{p}$
in the context of group entropy approach.
Now we explore some consequences about this by
considering the $G$-deformed Hamiltonian
$\hat{H}_G$
\begin{eqnarray}\label{G-quantum-Hamiltonian}
\hat{H}_G=\frac{1}{2m_0}\hat{p}_G^2 + V(\hat{x})=
%\nonumber\\
%&
\frac{1}{2m_0}\left[A(\hat{x})\hat{p}/2+
\hat{p}A(\hat{x})/2\right]^2
+ V(\hat{x})
\end{eqnarray}
of a single particle of mass $m_0$ having
%a $G$-deformed position $\hat{x}_G$ and
a $G$-deformed $\hat{p}_G$
and subjected to a potential $V(\hat{x})$,
being $A(z)=G^{\prime}(G^{-1}(z))$.
%From \eqref{G-quantum-Hamiltonian}, the eigenvalues equation
%$\hat{H}_G |\psi\rangle=E|\psi\rangle$
%for
%the autofunctions $\psi(x)=\langle x|\psi\rangle$ reads
%\begin{eqnarray}\label{G-Schrodinger-equation-1}
%\langle x |\hat{H}_G |\psi\rangle=E\psi(x).
%\end{eqnarray}
We also have the relationship
\begin{eqnarray}\label{G-momentum-operator-hermitic-commutator}
\hat{p}_G=A(\hat{x})\hat{p}-
\frac{1}{2}i\hbar A^{\prime}(\hat{x})
\end{eqnarray}
From this we have
\begin{eqnarray}\label{G-momentum-operator-hermitic-commutator2}
\hat{p}_G^2=[A(\hat{x})\hat{p}]^2
-\frac{1}{2}i\hbar\{A(\hat{x})\hat{p},A^{\prime}(\hat{x})\}
%\nonumber\\
%&
-\frac{1}{4}\hbar^2 [A^{\prime}(\hat{x})]^2
\end{eqnarray}
where $\{\hat{C},\hat{D}\}=\hat{C}\hat{D}+\hat{D}\hat{C}$
denotes the anti-commutator between $\hat{C}$ and $\hat{D}$.
%Since $\hat{p}=-i\hbar \partial/\partial_x$
Then, for an arbitrary
state $|\Psi\rangle$ we obtain
\begin{eqnarray}\label{G-momentum-matrix-elements-1}
&\langle x|\hat{p}_G^2|\Psi\rangle=
\langle x|[A(\hat{x})\hat{p}]^2|\Psi\rangle
-\frac{1}{2}i\hbar\langle x|
\{A(\hat{x})\hat{p},A^{\prime}(\hat{x})\}|\Psi\rangle
\nonumber\\
&
-\frac{1}{4}\hbar^2 (A^{\prime}(x))^2\langle x|\Psi\rangle=
A(x)^2\langle \hat{x}|\hat{p}^2|\Psi\rangle
-2i\hbar A(x)A^{\prime}(x)\langle \hat{x}|\hat{p}|\Psi\rangle \nonumber\\
&-\frac{1}{4}\hbar^2 [(A^{\prime}(x))^2+2A(x)A^{\prime\prime}(x)]
\langle x|\Psi\rangle.
\end{eqnarray}
Using that $\langle x^{\prime}|\hat{p}^n|\Psi\rangle=
(-i\hbar)^n\frac{\partial^n}{\partial x^n}\langle x^{\prime}|\Psi\rangle$
for all $n\in \mathbb{N}$ and $\langle x|\Psi\rangle=\Psi(x)$ we can recast the
Eq. \eqref{G-momentum-matrix-elements-1} to obtain
\begin{eqnarray}\label{G-momentum-matrix-elements-2}
\langle x|\hat{p}_G^2|\Psi\rangle=
&-\hbar^2A(x)^2 \frac{\partial^2 \Psi(x)}{\partial x^2}
-2\hbar^2A(x)A^{\prime}(x)\frac{\partial\Psi(x)}{\partial x}\nonumber\\
&-\frac{1}{4}\hbar^2 [A^{\prime}(x)^2+2A(x)A^{\prime\prime}(x)]\Psi(x).
\end{eqnarray}
From \eqref{G-quantum-Hamiltonian},
\eqref{G-momentum-matrix-elements-2} and
$i\hbar\frac{\partial}{\partial t}|\Psi(t)\rangle=
\hat{H}_G |\Psi(t)\rangle$ we arrive to
the
time-independent $G$-deformed Schr\"{o}dinger equation
for the wave-function $\Psi(x,t)$:
\begin{eqnarray}\label{G-deformed-SE}
&i\hbar \frac{\partial \Psi(x,t)}{\partial t}=
-\frac{\hbar^2}{2m_0}
A(x)^2 \frac{\partial^2 \Psi(x,t)}{\partial x^2}
-\frac{\hbar^2}{m_0} A(x)A^{\prime}(x)\frac{\partial\Psi(x,t)}{\partial x}\nonumber\\
&-\frac{1}{8m_0}\hbar^2 [A^{\prime}(x)^2+2A(x)A^{\prime\prime}(x)]\Psi(x,t)+
V(x)\Psi(x,t)
\end{eqnarray}
with $A(x)=G^{\prime}(G^{-1}(x))$ depending on the
group entropy $G$.
Several remarks deserve to be mentioned about the equation
\eqref{G-deformed-SE}.
First, the space deformation
$x_G=G^{-1}(x)=\int^{x} A(s)^{-1}ds$ implies that the particle has a
position-dependent
mass (that we call $G$-deformed mass $m(x)_G$)
\begin{eqnarray}\label{G-deformed-mass}
m(x)_G=m_0/A(x)^2=m_0/[G^{\prime}(G^{-1}(x))]^2
\end{eqnarray}
whose functional form depends exclusively of the
group entropy.
Second, the group entropy introduces
a new term of field
$-\frac{1}{8m_0}\hbar^2 [A^{\prime}(x)^2+2A(x)A^{\prime\prime}(x)]\Psi(x,t)$
depending on the derivatives of the deformation, which
physically can be interpreted as a field whose force
changes point by point due to the deformation.
And third, it is straightforwardly to show that
for the Tsallis class we have $A(x)=
G^{\prime}(G^{-1}(x))=(1+\gamma_qx)$ (with $\gamma_q=1-q$)
and then the $G$-deformed Schr\"{o}dinger equation \eqref{G-deformed-SE} results
\begin{eqnarray}\label{G-q-deformed-SE}
&i\hbar \frac{\partial \Psi(x,t)}{\partial t}=
-\frac{\hbar^2}{2m_0}
(1+\gamma_qx)^2 \frac{\partial^2 \Psi(x,t)}{\partial x^2}
-\frac{\hbar^2}{m_0}\gamma_q(1+\gamma_qx)\frac{\partial\Psi(x,t)}{\partial x}\nonumber\\
&-\frac{1}{8m_0}\hbar^2 \gamma_q^2\Psi(x,t)+
V(x)\Psi(x,t)
\end{eqnarray}
which is the $q$-deformed Schr\"{o}dinger equation corresponding
to a position-dependent mass $m(x)=m_0/(1+\gamma_qx)^2$,
presented previously in \cite{CostaFilho-Almeida-Farias-AndradeJr-2011,
Costa-Borges-2014,Costa-Borges-2018,Costa-Gomez-2018}.

Next step is to show that the $G$-deformed Schr\"{o}dinger
equation admits a rewrite in terms of the $G$-derivative
and having a constant mass. For accomplish this, we note that
the normalization condition for $\Psi(x,t)$ expressed with the
$G$-integral, i.e.
\begin{eqnarray}\label{G-normalization}
&1=\int \Psi(x,t)\Psi(x,t)^{*}dx=
\int_{(G)}\Psi(x,t)\Psi(x,t)^{*}dx_G(dx_G/dx)^{-1}\nonumber\\
& =\int_{(G)}\Phi_G(x,t)\Phi_G(x,t)^{*}dx_G
\end{eqnarray}
allows to identify the $G$-deformed wave-function
\begin{eqnarray}\label{G-deformed-wavefunction}
&\Phi_G(x,t)=\Psi(x,t)(\sqrt{dx_G/dx})^{-1}=\Psi(x,t)\sqrt{A(x)}\nonumber\\
&=\Psi(x,t)\sqrt{G^{\prime}(G^{-1}(x))}.
\end{eqnarray}
The $G$-deformed derivative $D_G=A(x)\frac{\partial}{\partial_x}$ applied two times
to $\Phi_G(x,t)$ gives
\begin{eqnarray}\label{G-derivative-SE-1}
&D_G^2 \Phi_G(x,t)=
\left(A(x)\frac{\partial}{\partial_x}\right)
\left(A(x)\frac{\partial \Phi_G(x,t)}{\partial_x}\right)=
A(x)^2\frac{\partial^2 \Phi_G(x,t)}{\partial x^2} \nonumber\\
&+A(x)A'(x)\frac{\partial \Phi_G(x,t)}{\partial x}
\end{eqnarray}
with
\begin{eqnarray}\label{G-derivative-SE-2}
\frac{\partial \Phi_G(x,t)}{\partial x}=
\sqrt{A(x)}\frac{\partial \Psi(x,t)}{\partial x}+\frac{A'(x)}{2\sqrt{A(x)}}\Psi(x,t)
\end{eqnarray}
and then
\begin{eqnarray}\label{G-derivative-SE-3}
\frac{\partial^2 \Phi_G(x,t)}{\partial x^2}=
\sqrt{A(x)}\frac{\partial^2 \Psi(x,t)}{\partial x^2}+
\frac{A'(x)}{\sqrt{A(x)}}\frac{\partial \Psi(x,t)}{\partial x}+\nonumber\\
 \left(\frac{2A''(x)\sqrt{A(x)}-A'(x)^2/\sqrt{A(x)}}{4 A(x)}\right)\Psi(x,t).
\end{eqnarray}
Thus, using Eqns.
\eqref{G-derivative-SE-2} and
\eqref{G-derivative-SE-3} we can recast
\eqref{G-derivative-SE-1} as
\begin{eqnarray}\label{G-derivative-SE-4}
D_G^2 \Phi_G(x,t)=\nonumber\\
\sqrt{A(x)}[A(x)^2\frac{\partial^2 \Psi(x,t)}{\partial x^2}+
2A(x)A'(x)\frac{\partial \Psi(x,t)}{\partial x}\nonumber\\
+\frac{1}{4}(A'(x)^2+2A(x)A''(x))\Psi(x,t)]
\end{eqnarray}
Therefore, multiplying the $G$-deformed
Schr\"{o}dinger equation \eqref{G-deformed-SE}
by $\sqrt{A(x)}$, using
\eqref{G-derivative-SE-4} and $x=x(x_G)$ we obtain
\begin{eqnarray}\label{G-SE}
i\hbar\frac{\partial \Phi_G(x_G,t)}{\partial t}=
-\frac{\hbar^2}{2m_0}D_G^2 \Phi_G(x_G,t)+
V(x_G)\Phi_G(x_G,t)
\end{eqnarray}
which is entirely written in the $G$-deformed
position coordinate $x_G=G^{-1}(x)$ and provided with $G$-deformed derivatives and a constant mass $m_0$.
It is clear that when there is no deformation ($G^{-1}(x)=x$), from Eq. \eqref{G-SE} we
recover the standard Schr\"{o}dinger equation, and for $G^{-1}(x)=x_q$ we obtain the $q$-deformed
Schr\"{o}dinger equation \eqref{G-q-deformed-SE} written in the $q$-deformed space.
In order to solve the Eq. \eqref{G-SE} we use the following properties
of the $G$-exponential:
\begin{eqnarray}\label{G-differential-equation}
D_G \exp_G(x)=G'(G^{-1}(x))\frac{\partial}{\partial x}\exp(G^{-1}(x))=
\exp_G(x_G)
\end{eqnarray}
where we have also
\begin{eqnarray}\label{G-derivative-G-number}
D_G f(x_G)=f'(x_G).
\end{eqnarray}
for all differential function $f(x)$.
From the group exponential $\exp_G(x)=\exp(x_G)$ we define the group cosine
and group sine as
\begin{eqnarray}\label{G-cosine}
\cos_G(x)=\cos(G^{-1}(x))=\cos(x_G)
\end{eqnarray}
and
\begin{eqnarray}\label{G-sine}
\sin_G(x)=\sin(G^{-1}(x))=\sin(x_G).
\end{eqnarray}
Using Eqns. \eqref{G-cosine}, \eqref{G-sine}
we obtain a $G$-deformed version of the De Moivre formula
\begin{eqnarray}\label{G-DeMoivre}
\exp(ix_G)=\cos_G(x)+i\sin_G(x)
\end{eqnarray}
Thus, the solution of the differential equation
\begin{eqnarray}\label{G-Helmoltz-equation}
D_G^2 \phi(x_G) + \alpha^2 \phi(x_G)=0
\end{eqnarray}
is $\phi(x_G)=A_1 \cos(\alpha x_G)+A_2 \sin(\alpha x_G)$
where $A_1,A_2$ are fixed numbers
determined by the initial conditions
$\phi(x_G)|_{x_G=x_G^{(0)}},D_G\phi(x_G)|_{x_G=x_G^{(0)}}$.

\subsection{Infinite potential well in the $G$-deformed space}

For illustrating the physical consequences of
the $G$-deformed
Schr\"{o}dinger equation we consider a
particle provided with a
position-dependent mass $m(x)_G=m_0/[G'(G^{-1}(x))]$
in a one-dimensional infinite potential well $V(x)$ of width $L$,
given by
\begin{equation}\label{potential}
V(x) = \left\lbrace
\begin{array}{ll}
0      &  \textup{iff} \quad 0\leq x\leq L \\
\infty &  \textup{iff} \quad x>L.
\end{array}
\right.
\end{equation}
Now if we assume that the deformation $x_G=G^{-1}(x)$ is an increasing function
of $x$ (this is the case of the Tsallis and the Kaniadakis classes)
then the potential well has the following expression in the
$G$-deformed space
\begin{equation}\label{G-potential}
V_G(x_G) = \left\lbrace
\begin{array}{ll}
0      &  \textup{iff} \quad 0\leq x_G\leq L_G \\
\infty &  \textup{iff} \quad x_G>L_G
\end{array}
\right.
\end{equation}
since $G(0)=0$ for all group entropy $G$ and $L_G=G^{-1}(L)$.
Physically, this means that the problem of a particle of a non-constant
mass $m_G(x)=m_0/[G'(G^{-1}(x))]^2$ in an infinite potential well of
length $L$ is equivalent to the
problem of a particle of constant mass $m_0$
in the $G$-deformed space subjected to the same potential but of width $L_G$.
For obtaining the autofunctions $\Phi_G(x_G)$ of this problem
we need to solve
the $G$-deformed eigenvalue equation
\begin{equation}\label{G-SE-potential-well}
-\frac{\hbar^2}{2m_0}D_G^2\Phi_G(x_G)+V_G \Phi_G(x_G)
=E_G\Phi_G(x_G)
\end{equation}
with the boundary conditions $\Phi_G(x_G=0)=\Phi_G(x_G=L_G)=0$.
Basing us on the standard case we have
\begin{equation}\label{G-autofunctions}
\Phi_G^{(n)}(x_G) = \left\lbrace
\begin{array}{ll}
A_G^{(n)}\sin(k_G^{(n)}x_G)      &  \textup{iff} \quad 0\leq x_G\leq L_G \\
0 &  \textup{iff} \quad x_G>L_G
\end{array}
\right.
\end{equation}
where the $G$-deformed eigenenergies $E_G^{(n)}$ are
\begin{eqnarray}\label{G-energies}
&E_G^{(n)}=\hbar^2 (k_G^{(n)})^2/2m_0, \nonumber\\
&\textrm{and}\nonumber\\
&k_G^{(n)}=n\pi/L_G=n\pi/G^{-1}(L) \quad \quad n=0,1,2,3,\ldots
\end{eqnarray}
being $(A_G^{(n)})^2=2/L_G$ the $G$-deformed normalization constant.
Now that we have solved the deformed Schr\"{o}dinger equation
\eqref{G-q-deformed-SE}
in the
$G$-deformed space, by means of the transformation
$x_G\mapsto x$
we can obtain the corresponding
autofunctions $\Psi_n(x)=\Phi_G^{(n)}(x)/\sqrt{G'(G^{-1}(x))}$
(see Eq. \eqref{G-deformed-wavefunction})
in the standard position space $x$:
\begin{equation}\label{deformed-autofunctions}
\Psi_n(x) = \left\lbrace
\begin{array}{ll}
\sqrt{\frac{2}{G^{-1}(L)G'(G^{-1}(x))}}\sin(n\pi G^{-1}(x)/G^{-1}(L)) &  \textup{iff} \quad 0\leq x\leq L \\
0 &  \textup{iff} \quad x>L.
\end{array}
\right.
\end{equation}
We can see how the formal group law $\oplus_G$
underlies the zeros spacing of $\Psi_n$.
From \eqref{deformed-autofunctions} for all $n\geq1$ the $n+1$ zeros of $\Psi_n$
are
\begin{equation}\label{deformed-zeros-1}
0, G\left(\frac{1}{n}G^{-1}(L)\right), G\left(\frac{2}{n}G^{-1}(L)\right),
\ldots, G\left(\frac{n-1}{n}G^{-1}(L)\right), L
\end{equation}
which obey precisely the formal group law ($G$-sum):
\begin{eqnarray}\label{deformed-zeros-2}
0, \oplus_G^{1}G(L_G/n), \oplus_G^{2}G(L_G/n), \ldots,
\oplus_G^{n-1}G(L_G/n), \oplus_G^{n}G(L_G/n)
\end{eqnarray}
It is instructive to illustrate the non-uniform zeros spacing for the Tsallis
and the Kaniadakis classes. If we call
\begin{equation}\label{G-spacing}
\Delta_G^{(m)}=G\left(\frac{m}{n}G^{-1}(L)\right)-G\left(\frac{m-1}{n}G^{-1}(L)\right)
\end{equation}
the $m$-th spacing\footnote{Which naturally reduces to
the uniform distribution of zeros
$\Delta^{(m)}=L/m$ for the standard case $G(t)=t$.} (with $1\leq m\leq n$)
then we have
\begin{equation}\label{G-spacing-Tsallis}
\Delta_q^{(m)}=\frac{1}{1-q}((1+(1-q)L)^{\frac{m}{n}}-(1+(1-q)L)^{\frac{m-1}{n}})
\end{equation}
for the Tsallis case, and
\begin{eqnarray}\label{G-spacing-Kaniadakis}
&\Delta_\kappa^{(m)}=\frac{1}{2\kappa}
\left((\kappa L+\sqrt{(\kappa L)^2+1})^{\frac{m}{n}}
+(\kappa L+\sqrt{(\kappa L)^2+1})^{\frac{m-1}{n}}\right) \nonumber\\
&-\frac{1}{2\kappa}
\left((\kappa L+\sqrt{(\kappa L)^2+1})^{-\frac{m}{n}}
-(\kappa L+\sqrt{(\kappa L)^2+1})^{-\frac{m-1}{n}}\right)
\end{eqnarray}
for the Kaniadakis case.
\begin{figure}[hbt]
\begin{minipage}[b]{1\linewidth}
\includegraphics[width=\linewidth]{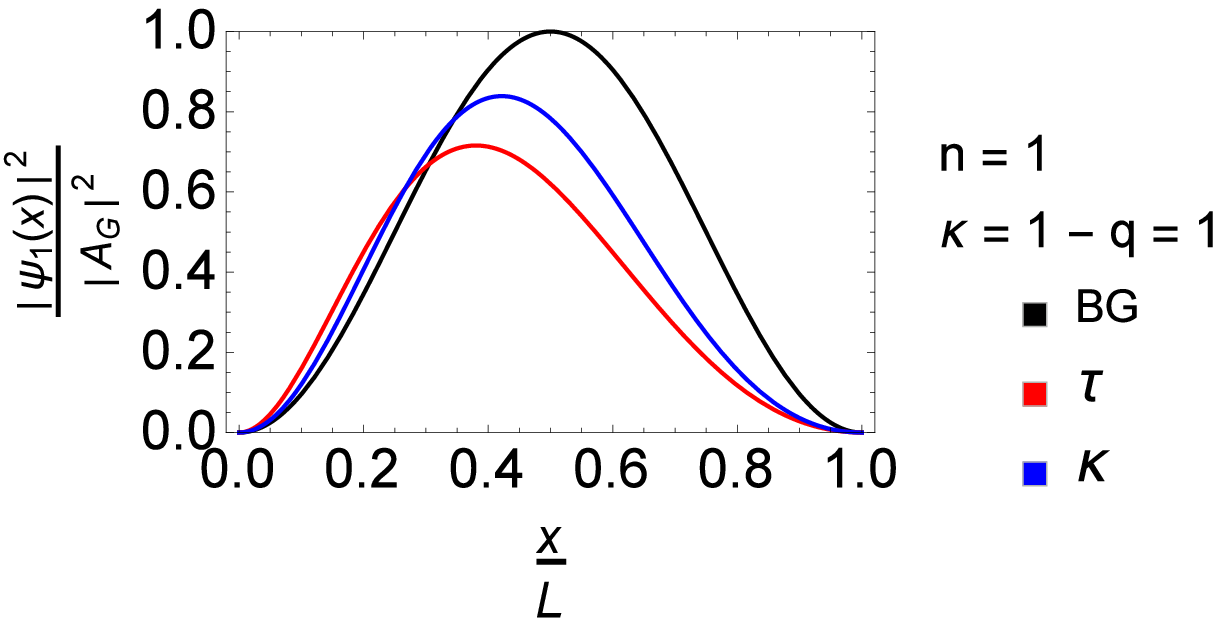}
\end{minipage}\\
\begin{minipage}[b]{1\linewidth}
\includegraphics[width=\linewidth]{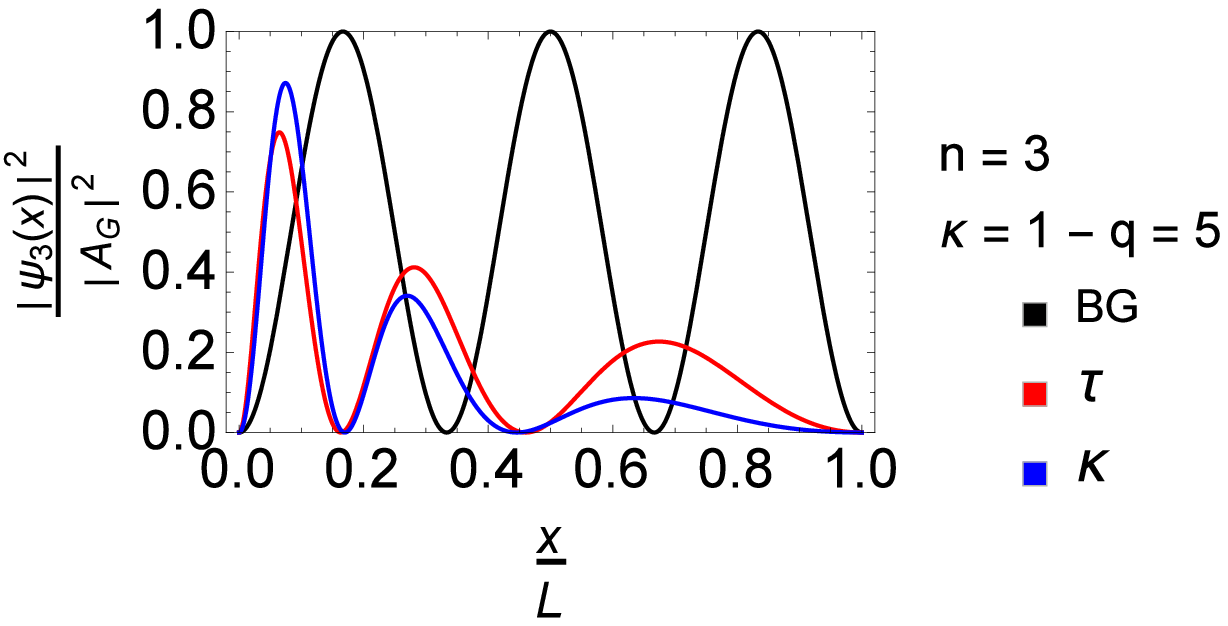}
\end{minipage}\\
\begin{minipage}[b]{1\linewidth}
\includegraphics[width=\linewidth]{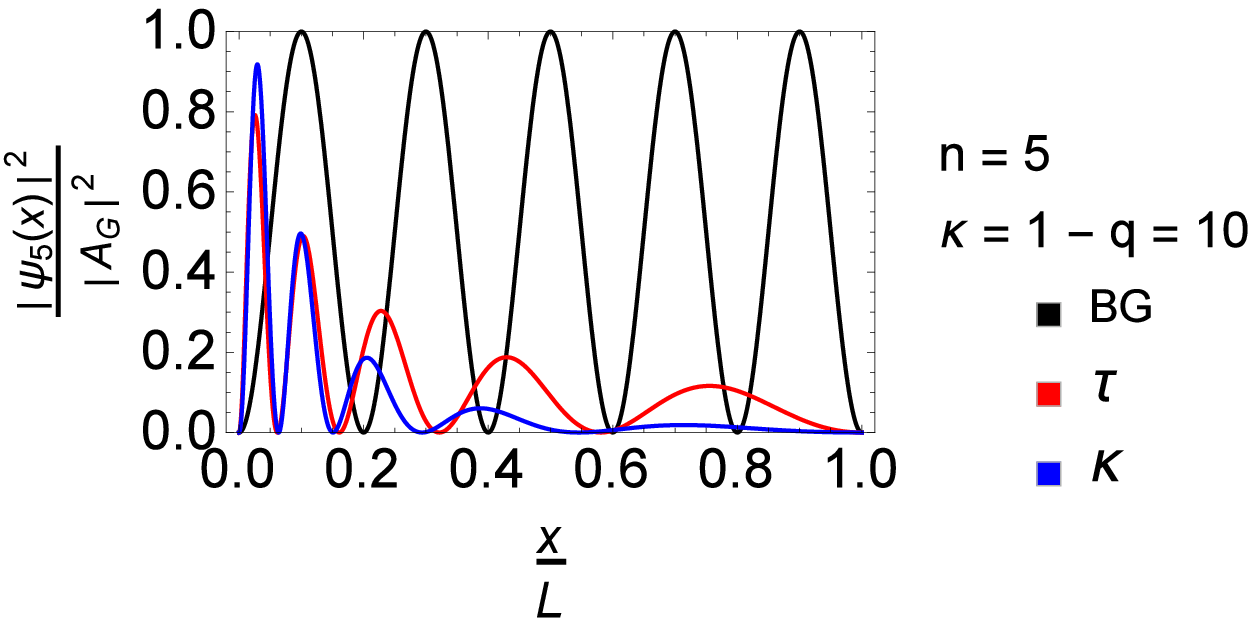}
\end{minipage}
\caption{\label{fig1}
Some plots of $|\Psi_n(x)|^2/|A|^2$ vs. $x/L$
where $\Psi_n(x)$ is the $n$-th autofunction
for the BG, Tsallis and Kappa classes and for some
values of $\kappa=1-q$.
The deformation breaks the symmetry of the autofunctions, by accumulating
probability density in the region where $m_G(x)$ (Eq. \eqref{deformed-autofunctions}) increases.
}
\end{figure}
In Fig. 1 we illustrate the probability distributions of some autofunctions
\eqref{deformed-autofunctions}
for the Boltzmann (standard case), Tsallis and Kaniadakis classes.
For comparing we set the $\kappa$ and $q$ parameters such that
$\kappa=1-q$. As the deformation increases
%($\kappa=1-q=10, 100$ in Fig. 1) this produces
a rapidly breaking
of the symmetry of the wavefunctions takes place
and the probability density tends to accumulate
in the zones where the mass takes big values, i.e. around
$x=0$. Thus, when the deformation is appreciable the particle
has a big mass around $x=0$ and then the probability to find it
near to $x=L$ is vanishingly small.
%On the other hand, if the deformation is low
%($\kappa=1-q=0.1; 1$ in Fig. 1) the probability is
%almost identical to the standard case. Moreover,
%the convergence of the Tsallis class to the Boltzmann one
%is more faster than the corresponding to the Kaniadakis class,
%as we can see for $n=3$ and $\kappa=1-q=0.1$, where
%the Tsallis class is completely superposed with the Boltzmann class while
%the Kaniadakis one still presents a dephasing.
Moreover, we observe
that the deformation of the probability distributions
is slightly more pronounced
for the Kaniadakis case than the Tsallis one.
\begin{figure}[hbt]
\begin{minipage}[b]{1\linewidth}
\includegraphics[width=\linewidth]{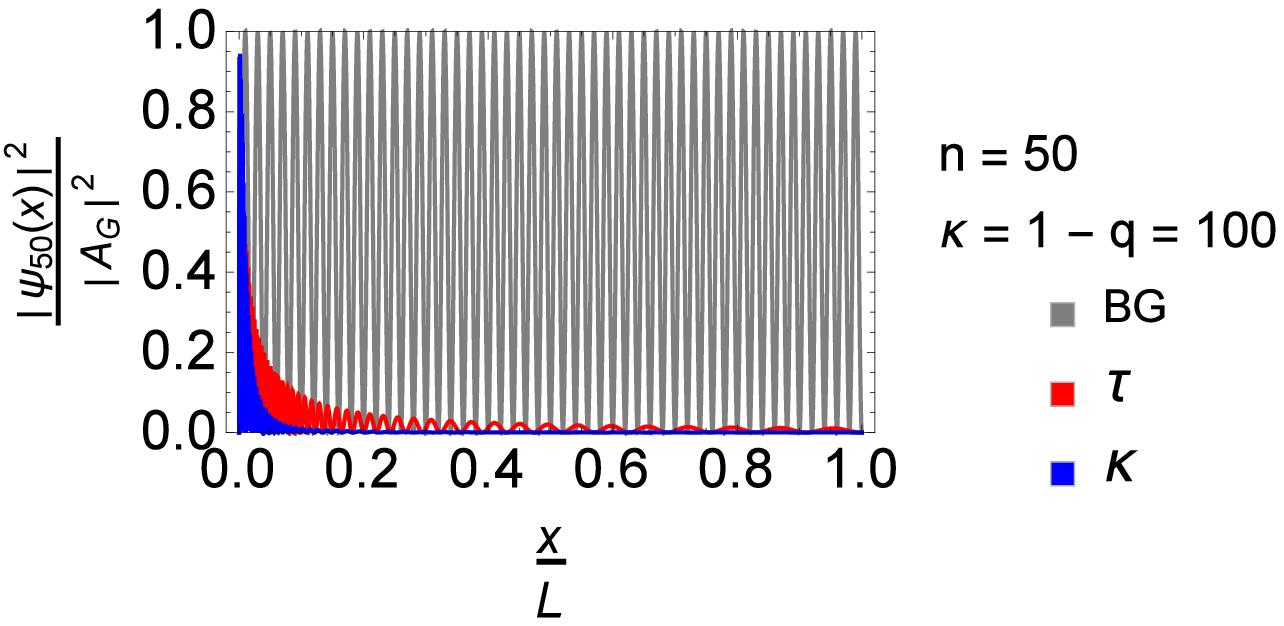}
\end{minipage}\\
\begin{minipage}[b]{1\linewidth}
\includegraphics[width=\linewidth]{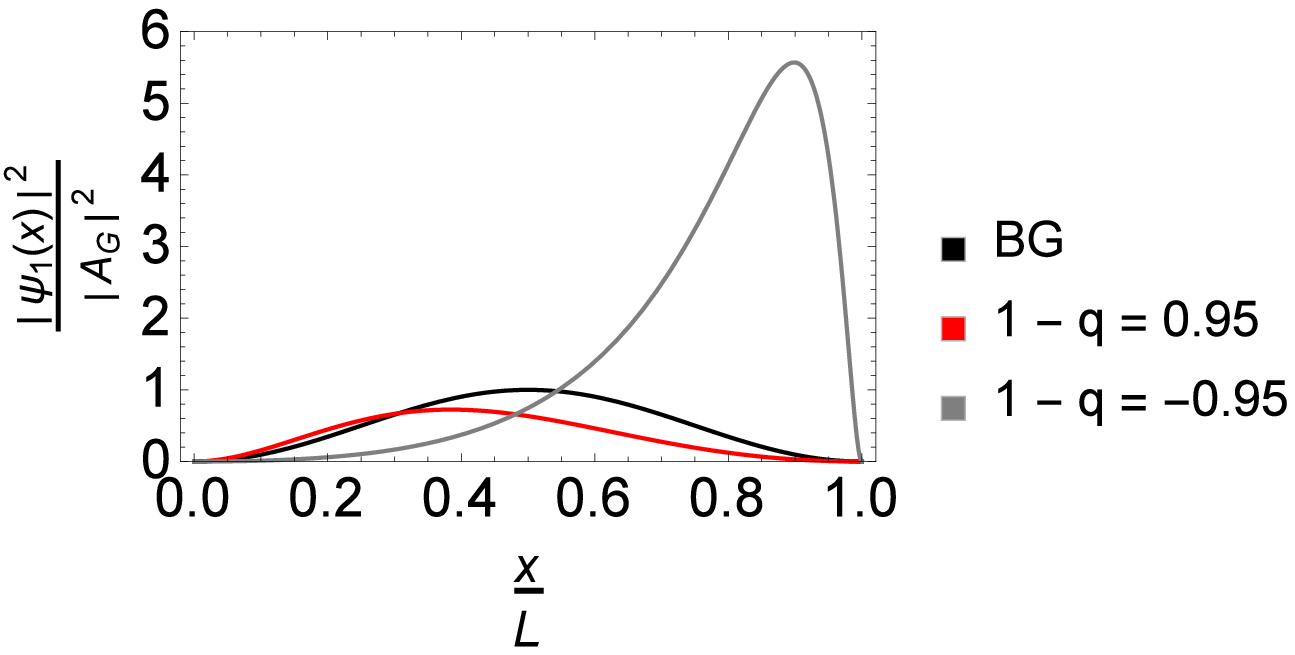}
\end{minipage}
\caption{\label{fig2}
Top panel:
the BG, Tsallis and Kappa classes for
$n=50$ and $\kappa=1-q=100$ are illustrated. The
deformed probability distributions tend
to a delta one around $x=0$ while the BG standard
one tends to the uniform distribution, as expected classically.
Bottom panel: for the fundamental state
we see that the deformation of the probability
distributions of the Tsallis class
are not symmetric with respect to the parameter $1-q$.
As $1-q$ approaches to $-1$ the deformed
probability distribution tends to a delta
around $x=1$.
%probability density in the region where $m_G(x)$ (Eq. \eqref{deformed-autofunctions}) increases.
}
\end{figure}

In the top panel of Fig. 2 we illustrate the classical limit ($n=50$) of the wavefunction probability
distributions
for the standard case and for the Tsallis and Kappa classes.
In both classes (Tsallis and Kappa ones)
the classical probability distribution tends to a delta probability distribution centered
at $x=0$ due to the particular form of the group entropies
$G_q$ and $G_\kappa$.
Regarding the parameter $1-q$, the asymmetry of
the deformed probability distribution of the Tsallis class
is depicted in the bottom panel of Fig. 2. We observe that as the parameter
$q$ tends to 2 the deformed distribution tends to accumulate
probability around $x=1$ due to the argument $(1+(1-q)x)$ makes to diverge
the distribution for $1-q=-1$.
By last, we can see that the similarity between the functional form
of the Tsallis mass
\begin{equation}\label{G-mass-Tsallis}
m_q(x)=\frac{m_0}{(1+(1-q)x)^2}
\end{equation}
and the Kaniadakis mass
\begin{equation}\label{G-mass-Tsallis}
m_\kappa(x)=\frac{m_0}{(\kappa x)^2+1},
\end{equation}
implies that they are essentially the same for $\kappa=1-q\gg 1$ because
$m_q(x)/m_\kappa(x)\sim (\kappa x)^2/(1-q)x^2=1$.
This is reflected in the fact of not observing a behavior substantially
different
between their corresponding autofunctions, for the example and the range
of parameters studied.

\section{Conclusions}\label{sec conclusions}

We have presented the algebraic structure (that we called
$G$-algebra) along with its associated
differential calculus, which is derived from the formal Lazard
group law and the exponential and logarithm groups.
From the $G$-algebra and its $G$-calculus
we have explored some of its consequences, that are summarized in Table 1.
\begin{table}[!htb]\label{tabla}
	\centering
	\begin{tabular}{|c||c|}
		\hline
		Structure & Group entropy structure
		\\
		\hline
		\hline
		algebra of real numbers & $G$-algebra
		\\
        \hline
        integers $n\in\mathbb{Z}$ & $G$-integers: $n_G=G(nG(1))$
        \\
        \hline
        real number $x$ & $G$-real number: $x_G=G^{-1}(x)$
        \\
        \hline
        differential $dx$ & $G$-differential: $dx_G=\frac{dx}{G^{\prime}(G^{-1}(x))}$
        \\
        \hline
		derivative $\frac{\partial}{\partial_x}$ & $G$-derivative $D_G$: $G^{\prime}(G^{-1}(x))\frac{\partial}{\partial_x}$
		\\
		\hline
		integral $\int f(x)dx $&  $G$-integral: $\int_{(G)}f(x_G)dx_G$
		\\
		\hline
                                                 & $G$-deformed operators $\hat{x}_G,\hat{p}_G$
        \\
        conjugated operators $\hat{x},\hat{p}$   & $\hat{x}_G=G^{-1}(\hat{x})$
        \\
                                                 & $\hat{p}_G=\frac{1}{2}\{G^{\prime}(G^{-1}(\hat{x})),\hat{p}\}$
        \\
        $[\hat{x},\hat{p}]=i\hbar\hat{1}$        &   $[\hat{x}_G,\hat{p}_G]=i\hbar\hat{1}$
        \\
        \hline
        Schr\"{o}dinger equation            & $G$-Schr\"{o}dinger equation
        \\
        $i\hbar\frac{\partial \Psi}{\partial t}
        =-\frac{\hbar^2}{2m_0}\frac{\partial^2\Psi}{\partial x^2}+V\Psi$  &         $i\hbar\frac{\partial \Phi_G}{\partial t}
        =-\frac{\hbar^2}{2m_0}D_G^2\Phi_G+V\Phi_G$
        \\
        \hline
        wavefunction           & $G$-wavefunction:
        \\
        $\Psi(x,t)$             &    $\Phi_G(x,t)=\Psi(x,t)\sqrt{G^{\prime}(G^{-1}(x))}$
        \\
        \hline
	\end{tabular}
	\caption{Several structures derived from the group entropy approach, by means of the
            formal group law, are listed. The application to the quantum mechanics
            implies the canonical transformation
            $\{\hat{x},\hat{p}\}\mapsto\{\hat{x}_G,\hat{p}_G\}$
                where the form of the $G$-Schr\"{o}dinger equation remains invariant.}
\end{table}

We point out the following consequences of our contribution.

First, we have obtained the $q$-algebra and the
$\kappa$-algebra as special cases for the Tsallis and
Kaniadakis group entropies respectively. Also, their
associated calculus, constituted by the
deformed derivative and integral, have been consistently derived.
Thus, for each choice of the group entropy one can obtain
its corresponding algebra and calculus by means of the $G$-algebra
and $G$-calculus formulas.

Second, we have given some clues in the connection of
the non-relativistic quantum mechanics and the entropy group approach,
by proposing a $G$-deformed Schr\"{o}dinger equation that
can derived from the $G$-calculus in a deformed position space.
We have shown that the $G$-deformed position and momentum operators
constitute a canonical transformation of the standard ones
and therefore, the motion equations
have as those of standard (non-deformed) quantum mechanics.
As a consequence, we have obtained
the Schr\"{o}dinger equation derived from the $G$-deformed position and momentum operators
(that we called $G$-deformed Schr\"{o}dinger equation), which
is physically identical to the corresponding of a particle having
a position-dependent mass, where the functional form of the mass
is univocally determined by the derivatives of the group entropy.
Furthermore, from the rewriting of the $G$-deformed Schr\"{o}dinger in terms
of the $G$-derivative and position and provided with a constant mass,
we have shown the equivalence between the problem of a particle having
an arbitrary position-dependent mass and the corresponding to a
particle of constant mass
within a $G$-deformed space. This connection constitutes an extension
of the investigated in previous works,
but for any arbitrary form of the functional mass.

Third, by studying the Tsallis and Kaniadakis classes
applied to the unidimensional motion
of a particle in a infinite potential well,
we have characterized the effect of the deformation of both group
entropies.
As a consequence of the deformation, the probability distributions of the Tsallis and
Kappa classes present an asymmetry which is more pronounced as the parameter
$\kappa=1-q$ increases (see Fig. 1).
We have found that the deformation does not affect the classical limit but
rather the way in which the probability density is distributed along the position space
(top panel of Fig. 2).
Due to the particular forms of the Tsallis and Kaniadakis mass functionals,
the probability density concentrates mostly in the region where the corresponding
mass functionals have big values (near to the $x=0$), as physically expected.

Finally, we consider that the results of the present work can be useful
for extending the group entropy approach
in other directions as diffusion phenomena, field theory, etc.
and also for complementing others recently obtained \cite{Tem19}.

\section*{Acknowledgments}
ISG and EPB acknowledge support received from the National Institute of Science
and Technology for Complex Systems (INCT-SC),
and from the National Council for Scientific and Technological Development
(CNPq) (at Universidade Federal da Bahia), Brazil.

\end{document}